\documentclass[12pt,preprint]{aastex}

\begin{document}

\title{Does The Addition of a Duration Improve the $L_{iso} - E_{peak}$ Relation For Gamma-Ray Bursts?}
\author{Andrew C. Collazzi and Bradley E. Schaefer}\affil{Physics and Astronomy, Louisiana State University, Baton Rouge, LA 70803}

\begin{abstract}

Firmani et al. proposed a new Gamma Ray Burst (GRB) luminosity relation that showed a significant improvement over the $L_{iso} - E_{peak}$ relation.  ($L_{iso}$ is the isotropic peak luminosity and $E_{peak}$ is the photon energy of the spectral peak for the burst.) The new proposed relation simply modifies the $E_{peak}$ value by multiplying it by a power of $T_{0.45}$, where $T_{0.45}$ is a particular measure of the GRB duration.  We begin by reproducing the results of Firmani for his 19 bursts.  We then test the Firmani relation for the same 19 bursts except that we use independently measured values for $L_{iso}$, $T_{0.45}$, and $E_{peak}$, and we find that the relation deteriorates substantially.  We further test the relation by using 60 GRBs with measured spectroscopic redshifts, and find a relation that has a comparable scatter as the original $L_{iso} - E_{peak}$ relation.  That is, a much larger sample of bursts does not reproduce the small scatter as reported by Firmani et al.  Finally, we investigate whether the Firmani relation is improved by the use of any of 32 measures of duration (e.g., $T_{90}$, $T_{50}$, $T_{90}/N_{peak}$, the fluence divided by the peak flux, $T_{0.30}$, and $T_{0.60}$) in place of $T_{0.45}$.  The quality of each alternative duration measure is evaluated with the root mean square of the scatter between the observed and fitted logarithmic $L_{iso}$ values.  Although we find some durations yield slightly better results than $T_{0.45}$, the differences between the duration measures are minimal. We find that the addition of a duration does not add any significant improvement to the $L_{iso} - E_{peak}$ relation.  We also present a simple and direct derivation of the Firmani relation from {\it both} the $L_{iso} - E_{peak}$ and Amati relations.  In all we conclude that the Firmani relation neither has an independent existence nor does it provide any significant improvement on previously known relations that are simpler.

\end{abstract}

\section{Introduction}

To date, eight separate luminosity relations have been identified for long duration Gamma-Ray Bursts (GRBs) (Schaefer and Collazzi, 2007). These relations correlate a burst's peak bolometric luminosity with various light curve and spectral parameters. The possible utility of these relations is that we can then use GRBs as tracers of the high-redshift universe.

One of these eight relations is the $L_{iso} - E_{peak}T_{0.45}$ relation proposed by Firmani et al. (2006, hereafter named the Firmani relation). Here, $E_{peak}$ describes the peak of the $E*F(E)$ curve (proportional to $\nu F_{\nu}$ ), which is the photon energy of the peak spectral flux; $L_{iso}$ is the isotropic luminosity of the burst measured bolometrically (1-10,000 kev in the burst rest frame). $T_{0.45}$ is the Reichart definition of a GRB time duration (Reichart et al. 2001) where the duration is the total time interval of the brightest bins in the light curve that contains 45$\%$ of the burst fluence. The Firmani relation was presented as an improvement over the $L_{iso} - E_{peak}$ relation. Nineteen GRBs were used to demonstrate a tight correlation with a reduced chi-square of 0.7 over 16 degrees of freedom; the resulting luminosity relation being $L_{iso} \propto {E_{peak}}^{1.62} {T_{0.45}}^{-0.49}$.  The reported scatter in the Firmani relation is substantially smaller than those of most other GRB luminosity relations.  This result offers the hope of substantial improvement in the accuracy of GRBs for cosmological distance measures.

We have no physical reason to expect that the addition of a duration should make for a tighter relation. Nonetheless, we should look to see if we can get tighter luminosity relations from duration definitions other than $T_{0.45}$, as $T_{0.45}$ may not be the optimal duration to use.  We have no physical reason to expect that any one definition of duration is best, while the particular choice of the Reichart definition was used only for historical reasons no longer of any relevance.  For example, we could still use the Reichart definition, but measure the duration over a different percentage of the burst fluence.  So perhaps the use of a duration based on 30\% or 60\% ($T_{0.30}$ or $T_{0.60}$) might be better.  Alternative definitions of duration can be considered instead of the Reichart formulation. For example, we can try to define the duration as equalling the total fluence divided by the peak flux to get a sort of equivalent width; alternatively, we could use the familiar $T_{90}$ or $T_{50}$ durations.  A wide variety of alternative durations can be defined, and we do not know which one will be optimal.

In this Section 2, we first reproduce the Firmani relation for the original data of Firmani et al. (2006) as a test that we are using identical fitting procedures. Then we test the Firmani relation with a set of independent data for the same 19 bursts.  A further test of the Firmani relation is made with a much larger sample of 60 bursts. In Section 3, we test many duration definitions in a Firmani-like relation to see which produces the `tightest' correlation. Section 4 contains a simple and forced derivation of the Firmani relation from two other luminosity relations.

\section{Testing The Firmani Relation}

We first started with the model as stated in Firmani et al. (2006):
\begin{center}
\begin{equation}
L_{iso} = 10^{52.11\pm0.03}\left(\frac{E_{peak}}{10^{2.37} keV}\right)^{1.62\pm0.08}\left(\frac{T_{0.45}}{10^{0.46} s}\right)^{-0.49\pm0.07} \frac{erg}{s}.
\end{equation}
\end{center}
We generalized this equation, and put it into logarithmic form:
\begin{center}
\begin{equation}
\log_{10}(L_{iso}) = \gamma + \xi *\log_{10}\left(\frac{E_{peak}}{10^{2.37} keV}\right) + \eta *\log_{10}\left(\frac{T_{0.45}}{10^{0.46} s}\right).
\end{equation}
\end{center}
Here $\gamma$, $\xi$, and $\eta$ are the fit parameters derived from fitting a set of GRB data, which can then predict model values of $\log_{10}(L_{iso})$. For the one-sigma uncertainty used in evaluating the chi-square, we used an elliptical error box to account for the errors in the measured quantities on both axes.  Specifically,
\begin{center}
\begin{equation}
\sigma_{combined}^2 = \sigma^2_{\log_{10}(L_{iso})} + \left(\frac{0.434 \xi \sigma_{E_{peak}}}{E_{peak}}\right)^2 + \left(\frac{0.434 \eta \sigma_{T_{0.45}}}{T_{0.45}}\right)^2.
\end{equation}
\end{center}
With the model values for the luminosity and its uncertainty as well as the observed luminosity (all in logarithmic space), we can calculate the reduced chi-square for the model fit.

To perform this test on the 19 bursts, we used data as reported in Firmani et al. (2006):  Their values for $T_{0.45}$ were the observed durations, and thus needed to be corrected for time dilation into the burst frame via dividing by $(1+\it{z})$.  Their values of $E_{peak}$ were already in the burst frame, and didn't need to be corrected.  $L_{iso}$ was obtained by taking their value of $L_{iso}/E_{iso}$ and multiplying it by their value $E_{iso}$.  We also used Firmani's reported redshift values.

Our best fit had a ${\chi_{r}}^2 = 0.7$ with an root-mean-square (RMS) scatter of the observed values of $\log_{10}(L_{iso})$ about the best fit model equal to $0.14$.  With 19 burst and 3 fit parameters, we have 16 degrees of freedom. This is in agreement with the reported value of $\chi^2_r = 0.7$ reported by Firmani et. al. (2006). We also agree with the best fit parameters and their uncertainties as reported by Firmani et al. (2006) and as given in Eq. 1. Thus, we are able to reproduce their reported result. We will use these procedures for all subsequent fits. Figure 1 shows the Firmani relation as we have reproduced it.

Given the nature of observational data, there are inevitably differences in the various published values of all these burst properties. For instance, Firmani et al. (2006) report $E_{peak}$= 685 $\pm$ 133 keV for GRB971214 , while Jimenez et al. (2001) reports a value of 840 $\pm$ 88 keV. For the same burst, the peak flux is slightly different for different detectors, so Fimani et al. (2006) report $\log_{10}(L_{iso})$ equals 52.86 $\pm$ 0.08, while we derive a value of 52.92 $\pm$ 0.01 (Schaefer 2007).  The Firmani relation should be robust on the use of independent measures from different published sources, so we should be able to obtain the same result with these different reported values.  For this test, we have collected independent measures for the luminosity, peak energy, and $T_{0.45}$ from various published reports.  We have no reason to think that either set of values for these 19 GRBs is better or worse in accuracy.  The independent values for these burst qualities are in Table 1.  Column 1 is the GRB designation; column 2 is the redshift; column 3 is the log of the isotropic bolometric luminosity; column 4 is the photon energy of the observed peak spectral energy (which needed to be blueshifted into the bursts' frame); and column 5 is the Reichart duration $T_{0.45}$.  The values and their references for the luminosities, peak energies, and redshifts can be found in Schaefer (2007).  We measured the durations for $T_{0.45}$ from the light curves referenced in Schaefer (2007).

The formal measurement error bars quoted for the $T_{0.45}$ values in Table 1 are fairly small, but the real total uncertainties are substantially larger.  We know this because various groups have reported measures of $T_{0.45}$ for many of the same GRBs and the scatter is much larger than anyone's quoted error bars.  In particular, we have collected independent measures reported by Guidorzi (2005), Firmani et al. (2006), Rizzuto et al. (2007), Rossi et al. (2008), Table 1 of this paper, and independent values calculated from our own group, with an average of five values for each of our bursts in Table 1.  We find that the median scatter is 17\%.  This value changes little if we look at bursts measured with one satellite ({\it Swift}) alone or if we only look at results from our group.  We think that this variation arises from relatively small changes resulting from differing time bin sizes and time intervals for the calculation.  (Similar scatter is found for other duration measures, see for example Koshut et al. 1996 and Norris et al. 1995.)  This additional systematic error contributes a small fraction of the extra scatter that we observe for the Firmani relation.  The reason for this small contribution is that the extra systematic error on $T_{0.45}$ is $\sim$12\% and the values are included nearly as a square root (see Eq. 1), so the extra contribution to $\sigma_{combined}$ is 0.026 (see Eq. 3).  All this is to say that the systematic errors in measuring $T_{0.45}$ are much larger than has ever been realized, yet even these additional uncertainties are negligibly small.

Upon optimizing the fit with the independent values, the equation for the best fit is:
\begin{center}
\begin{equation}
L_{iso} = 10^{52.09\pm0.02}\left(\frac{E_{peak}}{10^{2.37} keV}\right)^{1.90\pm0.05}\left(\frac{T_{0.45}}{10^{0.46} s}\right)^{-0.52\pm0.05} \frac{erg}{s}.
\end{equation}
\end{center}
A comparison with Eq. 1 shows that the two best fits are similar, with the exponent for the $E_{peak}$ being moderately different.  The Firmani relation for this independent data is displayed in Figure 2.  The obvious difference between Figures 1 and 2 is that Figure 2 has a much larger scatter than in Figure 1. The RMS value for the independent data was 0.35, whereas the RMS for the data from Firmani et al. (2006) is 0.14.  The reduced chi-square for the 19 bursts about this best fit model is  $\chi_{r}^{2} = 14.50$ for the independent data.  This is greatly larger than the value of $\chi_{r}^{2} = 0.7$ obtained from the data from Firmani et al. (2006).  With this large reduced chi-square, we realize that there must be some additional source of systematic uncertainty that is beyond that from ordinary measurement errors.

We can introduce an additional figure of merit which quantifies the scatter about the best fit Firmani relation.  This is the systematic error required to be added in quadrature to the measurement error such that the resulting reduced chi-square equals unity.  A desirable fit with little scatter will have a small required systematic contribution to the uncertainties, whereas a poor fit with large scatter will have a large required systematic contribution.  For this, we will assume that the systematic error is a constant, even though the reality is likely more complex in ways that we cannot now see.  In essence, we are calculating the uncertainty in the chi-square calculation as
\begin{center}
\begin{equation}
\sigma_{total}^2 = \sigma_{sys}^2 + \sigma_{combined}^2.
\end{equation}
\end{center}
For the case of Firmani's 19 GRBs with his data, there is no required additional systematic uncertainty (as indicated by the reduced chi-square being less than unity).  But for the independent data for the same bursts, we require a systematic error of 0.34 (in logarithmic units) to be added in quadrature so as to get an acceptable fit with a reduced chi-square of unity.  So in all, we can quantify the quality of the Firmani relation for any data set by three parameters, $\chi_r^2$, RMS, and $\sigma_{sys}$.  Table 2 summarizes these parameters for the Firmani relations with various data sets.

Our next step was to add more bursts. That is, the Firmani relation should be robust when applied to a much larger sample of bursts. Specifically, we used 60 of the bursts as given in Table 1, with the observed burst properties as collected in Schaefer (2007) based on references reported therein.  Schaefer (2007) tabulates 69 bursts in all, but several had to be omitted for various reasons. GRB 980613, GRB 990712, GRB 011211, and GRB 020903 were not used as we did not have the light curves for duration calculations. GRB 050824 was omitted because the value of $E_{peak}$ is only an upper limit. GRB 050319, GRB 050408, GRB 050802, and GRB 051111 were omitted due to the reported value of $E_{peak}$ in Schaefer 2007 not having been directly measured.

To remain consistent, we only used data reported in Schaefer (2007) for the peak energy and redshift. The values for both $L_{iso}$ and $\sigma_{L_{iso}}$ were derived from values for the bolometric peak flux ($P_{bolo}$) reported by Schaefer (2007).  With the standard inverse square law, we get
\begin{center}
\begin{equation}
L_{iso} = P_{bolo} *4 \pi d^2_L.
\end{equation}
\end{center}
The luminosity distance ($d_L$) to the GRB is calculated with the measured spectroscopic redshift, assuming the concordance cosmology ($\Omega_{M}$=0.27 in a flat universe with $w=-1$). 

With this independent data set for 60 GRBs, we fitted the model in Eq. 2.  The equation for the best fit for this extended sample is:
\begin{equation}
L_{iso} = 10^{52.09\pm0.01}\left(\frac{E_{peak}}{10^{2.37} keV}\right)^{1.91\pm0.03}\left(\frac{T_{0.45}}{10^{0.46} s}\right)^{-0.67\pm0.03} \frac{erg}{s}.
\end{equation}
This best fit model is similar to the best fits with the 19 GRB subsample (cf. Eqs 1 and 4).  The resulting Firmani relation is plotted in Figure 3.  Again, the immediate reaction is that the figure displays a lot of scatter, and much more scatter than in either Figures 1 or 2.  Quantitatively, the comparisons are presented in Table 2.  We see that the RMS scatter has risen to 0.41, which is greatly larger than in the earlier Figures.  The reduced chi-square of the fit is $\chi_r^2 = 15.89$, which shows that there is some source of scatter that is much larger than produced by the simple measurement uncertainties in the input parameters.  The systematic error for the 60 burst sample is 0.38, which is substantially larger than for either data set in the 19 burst subsample.

The primary result from this section is that the Firmani relation is neither robust to the use of independent data nor robust to the extension to many more bursts.

\section{Seeking the Optimal Duration}

Now that we have seen how the Firmani relation behaves for our sample of 60 GRBs, we should try the same test procedures for various different duration definitions.  For this, we use a generalized form of Eq. 2:
\begin{center}
\begin{equation}
\log_{10}(L_{iso}) = \gamma + \xi *\log_{10}\left(\frac{E_{peak}}{<E_{peak}>}\right) + \eta *\log_{10}\left(\frac{\tau}{<\tau>}\right).
\end{equation}
\end{center}
Here, the duration has been generically labeled as $\tau$, and the denominators inside the logarithms are constants equal to the average $\tau$ and $E_{peak}$ values for the data set. The reason to have these averages in the denominator is to improve the convergences of the fits by avoiding long thin error regions with strong correlations between fit parameters.

There are many alternative ways to measure duration. For example, we could use the Reichart definition, but with a different percentage of the total fluence to take the duration over. In other words, we could expand the Reichart definition of duration out to say $T_{0.60}$, or contract it to $T_{0.30}$. Again, there is no reason to believe that using the exact duration as proposed by Reichart ($T_{0.45}$) would be any more effective than the others.  Other duration definitions are reasonable, and indeed much easier to calculate.  We could adopt a duration defined as the time a burst spends above $x\%$ of the peak flux of the burst ($\mathcal{T}_{x}$).  The well-known definitions of duration, $T_{90}$ and $T_{50}$, should also be included. Here, the two durations are the time interval containing the central 90 or 50 percent of the fluence of the burst, respectively.  We also take the bolometric fluence $S_{bolo}$ and divide it by the bolometric peak flux $P_{bolo}$ to get a sort of `equivalent width'. 
 
As a control, we have also adopted a case where all the burst durations are set equal to a constant, which we arbitrarily take to be $\tau$=10 seconds.  The chosen value does not matter, as different choices will merely result in a different $\gamma$ value that will not change the quality of the fit.  By taking a constant duration, we are transforming the Firmani relation ($L_{iso} - E_{peak}T_{0.45}$) into the old $L_{iso} - E_{peak}$ relation.  A comparison of the scatter in the $\tau$=10 seconds relation versus the generalized Firmani relations will tell us whether the addition of a time scale has substantially improved the quality of the luminosity indicator.

So far, the alternative durations have all been measures of the total duration of the burst.  However, the physics of the luminosity relations points to the correlations as being with the individual peak pulse and not the overall set of pulses that make up the entire light curve (Schaefer 2003; 2004).  So we should here consider various pulse duration measures.  A simple and reasonable means of doing this is to take all of the overall-burst-duration measures and divide by the number of pulses in the light curve ($N_{peak}$, as defined in Schaefer 2007).  This immediately doubles the number of trial definitions considered.  We are left with a total of 32 duration definitions, all of which are listed in Table 3.

For all 60 bursts in our sample, we have measured the durations according to all 32 definitions.  We have calculated these durations from the light curves as given on the BATSE, HETE, {\it Konus}, and {\it Swift} websites as well as the values reported in Schaefer (2007).  For $T_{90}$ and $T_{50}$, we have almost always been able to find values reported by the instrument teams  (see Schaefer 2007 for references).
 
We fitted Eq. 8 for all 60 GRBs for all 32 duration measures.  For each best fit relation, we calculated the RMS and $\sigma_{sys}$ values as quantitative figures of merit. These are summarized in Table 3.

The results indicate that while there are certainly differences between duration definitions, the differences tend to be small.  The smallest values of RMS and $\sigma_{sys}$ occurs for durations defined as $\mathcal{T}_{30}/N_{peak}$.  The scatter in this best relation is somewhat smaller than for the original Firmani relation (with $T_{0.45}$ in the third line of Table 3).  However, we do not think that the differences are significant.  Our reason is that there will be noise in the figures of merit which will inevitably produce one duration definition as being the best even if the values were uncorrelated or random, and we see the scale for such variations by comparing the values in Table 2.  That is, the variations of the figures of merit in Table 3 are consistent with the case where duration information is not correlated with $L_{iso}$, and a different set of 60 GRBs would randomly produce a different 'best' definition.  As such, we do not think that we should replace the Firmani relation with a luminosity relation involving $\mathcal{T}_{30}/N_{peak}$.

A particularly important comparison is between the Firmani relation and the $L_{iso} - E_{peak}$ relation (represented by the line with the durations all taken to be a constant of 10 seconds).  We see that the $L_{iso} - E_{peak}$ relation is the third poorest relation in the table.  Nevertheless, the difference is not large enough to evaluate as being significant.  That is, the differences in the figures of merit (0.09 in the RMS and 0.08 in $\sigma_{sys}$) are too small to view as necessarily arising from a physical effect in the bursts.  We know this because the variation caused by simple sampling effects (see the last two lines of Table 2) are of order 0.06 in the RMS and 0.04 in $\sigma_{sys}$.  As such, we view the Firmani relation as having a similar scatter as the $L_{iso} - E_{peak}$ relation.

\section{Discussion}

In a recent independent preprint, Rossi et al. (2008) also examined the Firmani relation, in particular with a comparison to the Amati relation.  They use an extended sample of 40 {\it BeppoSAX} and {\it Swift} bursts, with little overlap with our sample of 60 GRBs.  Their best fit is somewhat different from those in Eqs 1, 4, or 7; with their fitted Firmani relation scaling as $L_{iso} \propto E_{peak}^2 T_{0.45}^{-1}$.  They realized that this Firmani relation is essentially identical to the Amati relation (Amati et al. 2002), which gives the isotropic energy emitted in gamma radiation over the whole burst duration as $E_{\gamma ,iso} \propto E_{peak}^2$.  With the reasonable approximation that the total energy in the light curve equals the peak luminosity times the duration ($E_{\gamma ,iso} \approx L_{iso} T_{0.45}$), the Amati relation ($E_{\gamma ,iso} \propto E_{peak}^2$) is transformed into their Firmani relation ($L_{iso} \propto E_{peak}^2 T_{0.45}^{-1}$).  While our exponents in the Firmani relation are somewhat different, the Rossi derivation demonstrates that  the Fermani relation has a physical basis that is close to that of the Amati relation.  Rossi et al. (2008) further go on to show that the scatter in their Firmani relation is comparable to that in the Amati relation, which is another way of saying that the two relations are not independent. 

Here, we will directly derive the Firmani relation from {\it both} the $L_{iso} - E_{peak}$ and Amati relations.  Let us start with the relation $L_{iso} \propto E_{peak}^{1.68}$ as given in Schaefer (2007).  This can be rearranged as $L_{iso} \propto E_{peak}^{1.9} (E_{peak}^2/L_{iso})^{-0.69}$.  The Amati relation ($E_{\gamma ,iso} \propto E_{peak}^2$) can be inserted to get $L_{iso} \propto E_{peak}^{1.9} (E_{\gamma ,iso}/L_{iso})^{-0.69}$.  Now we can select one of our duration definitions with $\tau = S_{bolo}/P_{bolo}$.  The ratio of fluence to peak flux will equal to the ratio of the burst energy and the peak luminosity, so we have $\tau = E_{\gamma ,iso}/L_{iso}$.  This can now be substituted to obtain $L_{iso} \propto E_{peak}^{1.9} \tau^{-0.69}$.  We see that we have just derived Eq. 8 with $\xi = 1.9$ and $\eta = -0.69$, values which are characteristic of the fitted Firmani relation (cf. Eq. 7).  With this, we see that the Firmani relation has no independent existence because it is only a combination of two simpler relations.

Thus, given any two of these three relations, we can derive the third. A question is which of these is more fundamental. The inherent problem with making this assessment is that it comes down to how one identifies the more fundamental of relations that really address different physics. By Occam's Razor, the more fundamental relation would be one that has the fewest parameters, while accurately and efficiently yielding a luminosity. Thus, the Firmani relation is less fundamental as it uses more parameters for the fit. Of the two relations that remain, the one that has the most `utility' will be the one with the least amount of scatter in its calibration curve.

We were able to reproduce Firmani's results over his small sample of 19 bursts. However, when we substitute independent values for $L_{iso}$, $E_{peak}$ and $T_{0.45}$, we see a substantial broadening around the model. That is, the Firmani relation is not robust on the use of alternative input data. In addition, when we extend the test to a larger sample of 60 bursts, the scatter becomes substantially larger again. Indeed, this scatter is comparable to the scatter in the original $E_{peak} - L_{iso}$ relation.  That is, the Firmani relation is not robust for the use of additional bursts.  These failures of the Firmani relation  have dashed our hopes raised by the tight calibration curves displayed in Firmani et al. (2006).  It also suggests that the addition of a duration does not significantly improve the $L_{iso} - E_{peak}$ relation.  The larger point of interest is that no duration shows a significant advantage over the $E_{peak} - L_{iso}$.  While it might be possible that that a relation involving $\mathcal{T}_{30}/N_{peak}$ might really have a smaller scatter than the Firmani relation, the improvements are small and not significant. This leads us to conclude that the addition of a duration is not doing enough to improve the $L_{iso} - E_{peak}$ relation to be considered to be a separate luminosity relation.

We conclude that the Firmani relation is not useful for several reasons:  First, the Firmani relation is simply derived by putting together two well-known, simpler, and independent luminosity relations, and thus it has no separate existence.   Second, it is not robust for the inclusion of independent input data or for the extension to many more GRBs.  Third, the real scatter for the Firmani relation does not live up to the hope generated by the original report, with the scatter being comparable to those of the luminosity relations from which it is derived.  In all, we can see no utility or advantage to using the Firmani relation.

We would like to thank the Louisiana State University Board of Regents and the Louisiana Space Consortium for their support.

{}
\clearpage

\begin{figure}[h!]
  \begin{center}
    \includegraphics[scale=1.0]{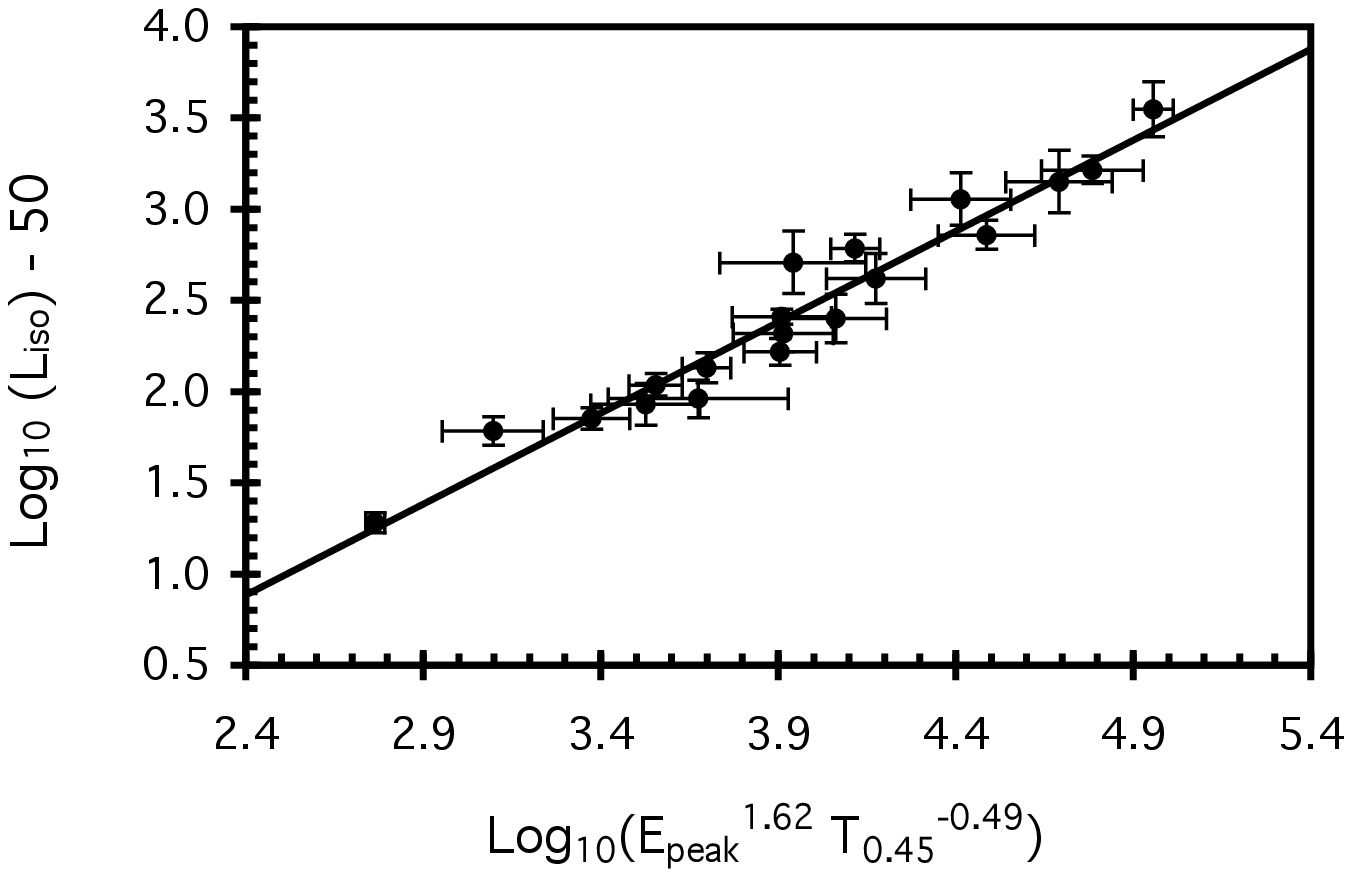}
  \end{center}
  \caption{The Firmani relation with Firmani's data. Here we have used all burst properties as reported by Firmani et al., (2006).  Because the scatter was so small, this result is potentially important as it would offer a means to substantially improve the calibration of distances to many GRBs.  Here, the y-axis is the logarithm of the luminosity which is in ergs per second. The subtraction of 50 is for easier comparison
to the graph as shown in Firmani et al., (2006). This y-axis convention is used on all subsequent plots for ease of comparing the results.}
\end{figure}

\clearpage

\begin{figure}[h!]
  \begin{center}
    \includegraphics[scale=1.0]{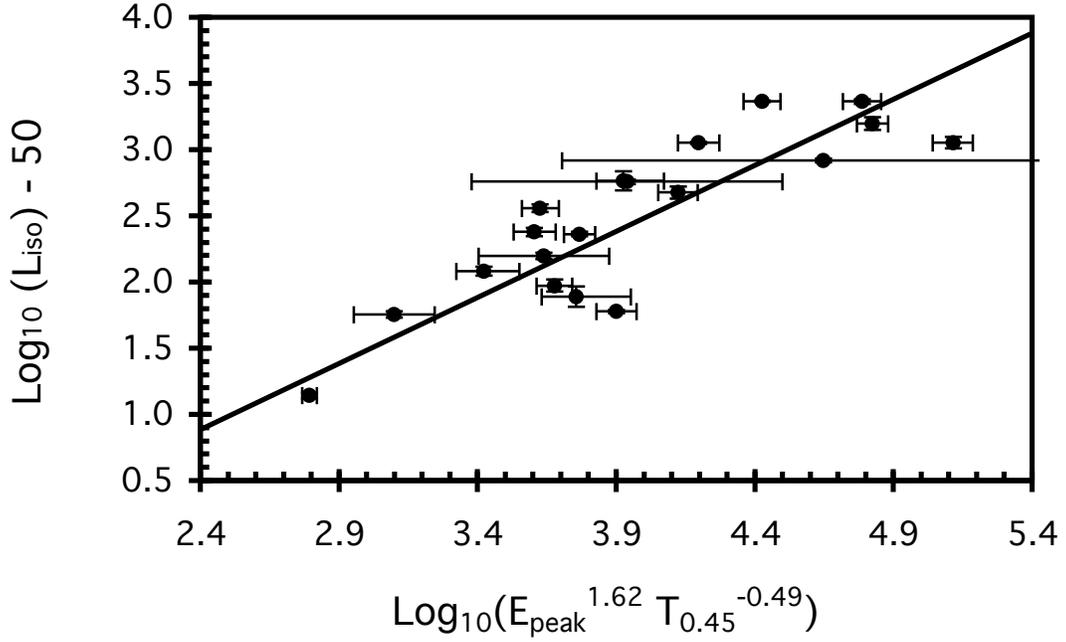}
  \end{center}
  \caption{The Firmani relation with independent data for the same 19 GRBs.  The main point from this figure is that the scatter is much greater than in Figure 1.  The chart area is matched to the previous one to make better comparison of the relations.  The line in this figure is identical with the line in Figure 1 (i.e., the original Firmani relation) as another aid for comparison.  The two best fits for the first two figures have slightly different exponents (see Eqs 1 and 4) so Figure 2 is slightly non-optimal in representing the best fit.}
\end{figure}

\clearpage

\begin{figure}[h!]
  \begin{center}
    \includegraphics[scale=1.0]{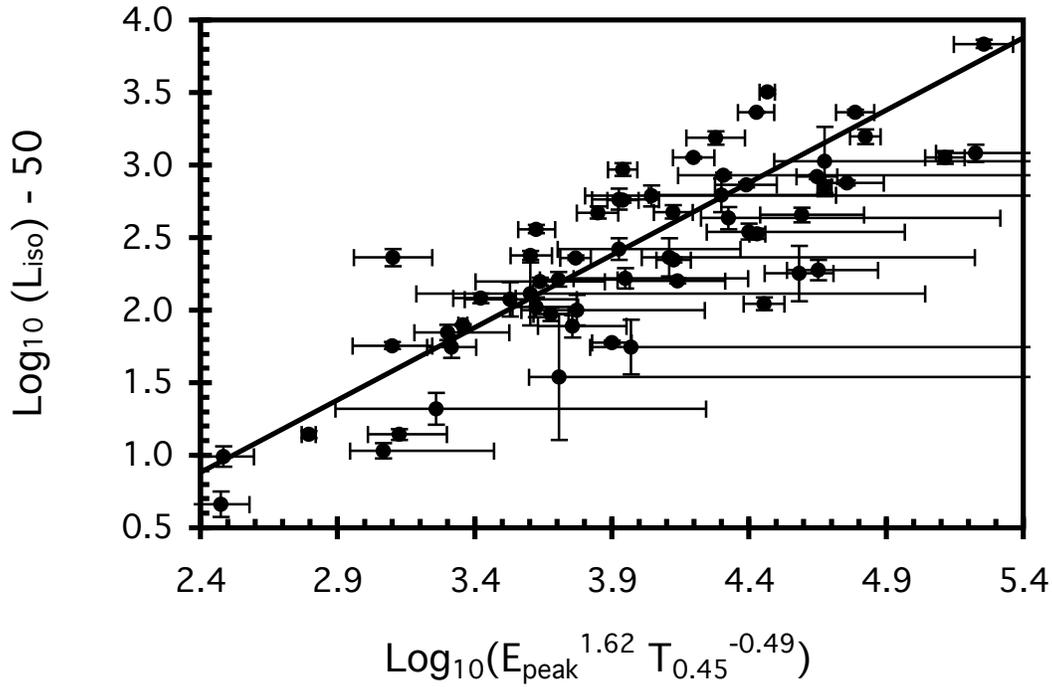}
  \end{center}
  \caption{The Firmani relation when extended to 60 GRBs.  The main point from this figure is that the scatter is much greater than in Figure 1, and is also significantly larger than in Figure 2.  This figure is given with identical axes and fit line (from Eq. 1) as the other figures to allow simple comparisons.  This scatter is comparable to that for the older $L_{iso}-E_{peak}$ relation, and this points to our main conclusion that the Firmani relation is not an improvement.}
\end{figure}

\clearpage

\begin{deluxetable}{ccccc}
\tabletypesize{\scriptsize}
\tablecolumns{5}
\tablewidth{0pc}
\tablecaption{Burst properties that were used throughout this paper. }
\tablehead{\colhead{BURST} &
\colhead{\textit{z}\tablenotemark{a}} & \colhead{$Log(L_{iso})$\tablenotemark{a}} & \colhead{$E_{peak}$\tablenotemark{a}\tablenotemark{f}} & \colhead{$T_{0.45}$\tablenotemark{b}}\\
& & (erg/s) & (keV) & (s) \\
(1) & (2) & (3) & (4) & (5) \\}
\startdata
970228\tablenotemark{c}	&	0.70	&	52.20	$\pm$	0.03	&	$115^{+38}_{-38}$  	&	2.37	$\pm$	0.44	\\
970508	&	0.84	&	52.04	$\pm$	0.04	&	$389^{+[40]}_{-[40]}  $  	&	4.03	$\pm$	0.24	\\
970828\tablenotemark{c}	&	0.96	&	52.68	$\pm$	0.05	&	$298^{+[30]}_{-[30]}  $  	&	10.50	$\pm$	0.45	\\
971214\tablenotemark{c}	&	3.42	&	52.92	$\pm$	0.01	&	$190^{+[20]}_{-[20]}  $  	&	6.72	$\pm$	0.09	\\
980703\tablenotemark{c}	&	0.97	&	51.78	$\pm$	0.01	&	$254^{+[25]}_{-[25]}  $  	&	18.00	$\pm$	1.80\tablenotemark{e}\\
990123\tablenotemark{c}	&	1.61	&	53.36	$\pm$	0.02	&	$604^{+[60]}_{-[60]}  $  	&	16.58	$\pm$	0.05	\\
990506\tablenotemark{c}	&	1.31	&	53.05	$\pm$	0.01	&	$283^{+[30]}_{-[30]}  $  	&	12.67	$\pm$	0.63\tablenotemark{d}	\\
990510\tablenotemark{c}	&	1.62	&	52.76	$\pm$	0.02	&	$126^{+[10]}_{-[10]}  $  	&	5.06	$\pm$	0.25\tablenotemark{d}	\\
990705\tablenotemark{c}	&	0.84	&	52.36	$\pm$	0.02	&	$189^{+15}_{-15}  $  	&	9.54	$\pm$	0.27	\\
991208	&	0.71	&	52.67	$\pm$	0.04	&	$190^{+[20]}_{-[20]}  $  	&	4.80	$\pm$	0.24\tablenotemark{d}	\\
991216\tablenotemark{c}	&	1.02	&	53.36	$\pm$	0.01	&	$318^{+[30]}_{-[30]}  $  	&	3.58	$\pm$	0.05	\\
000131\tablenotemark{c}	&	4.5	&	53.20	$\pm$	0.05	&	$163^{+13}_{-13}  $  	&	4.54	$\pm$	0.09	\\
000210	&	0.85	&	52.85	$\pm$	0.04	&	$408^{+14}_{-14}  $  	&	1.73	$\pm$	0.06	\\
000911\tablenotemark{c}	&	1.06	&	53.05	$\pm$	0.04	&	$986^{+[100]}_{-[100]}  $  	&	6.46	$\pm$	0.32\tablenotemark{d}	\\
000926	&	2.07	&	52.97	$\pm$	0.04	&	$100^{+7}_{-7}  $  	&	4.67	$\pm$	0.45	\\
010222	&	1.48	&	53.51	$\pm$	0.01	&	$309^{+12}_{-12}  $  	&	6.46	$\pm$	0.14	\\
010921	&	0.45	&	51.14	$\pm$	0.04	&	$89^{+21}_{-13.8}  $  	&	5.74	$\pm$	0.58	\\
020124\tablenotemark{c}	&	3.2	&	52.76	$\pm$	0.07	&	$87^{+18}_{-12}  $  	&	12.14	$\pm$	0.58	\\
020405	&	0.7	&	52.20	$\pm$	0.02	&	$364^{+90}_{-90}  $  	&	10.18	$\pm$	0.38	\\
020813\tablenotemark{c}	&	1.25	&	52.56	$\pm$	0.03	&	$142^{+14}_{-13}  $  	&	17.36	$\pm$	0.23	\\
021004	&	2.32	&	52.00	$\pm$	0.10	&	$80^{+53}_{-23}  $  	&	6.89	$\pm$	0.41	\\
021211\tablenotemark{c}	&	1.01	&	52.08	$\pm$	0.03	&	$46^{+8}_{-6}  $  	&	0.66	$\pm$	0.12	\\
030115	&	2.5	&	52.22	$\pm$	0.07	&	$83^{+53}_{-22}  $  	&	4.26	$\pm$	0.30	\\
030226\tablenotemark{c}	&	1.98	&	51.89	$\pm$	0.08	&	$97^{+27}_{-17}  $  	&	8.86	$\pm$	0.87	\\
030323	&	3.37	&	52.11	$\pm$	0.22	&	$44^{+90}_{-26}  $  	&	6.89	$\pm$	0.70	\\
030328\tablenotemark{c}	&	1.52	&	52.38	$\pm$	0.03	&	$126^{+14}_{-13}  $  	&	20.83	$\pm$	0.70	\\
030329\tablenotemark{c}	&	0.17	&	51.14	$\pm$	0.02	&	$68^{+2.3}_{-2.2}  $  	&	4.43	$\pm$	0.23	\\
030429	&	2.66	&	52.08	$\pm$	0.12	&	$35^{+12}_{-8}  $  	&	2.13	$\pm$	0.35	\\
030528	&	0.78	&	50.66	$\pm$	0.09	&	$32^{+4.7}_{-5}  $  	&	9.99	$\pm$	0.93	\\
040924\tablenotemark{c}	&	0.86	&	51.97	$\pm$	0.05	&	$67^{+6}_{-6}  $  	&	0.49	$\pm$	0.02\tablenotemark{d}	\\
041006\tablenotemark{c}	&	0.71	&	51.76	$\pm$	0.03	&	$63^{+13}_{-13}  $  	&	4.26	$\pm$	0.12	\\
050126	&	1.29	&	51.03	$\pm$	0.05	&	$47^{+27}_{-8}  $  	&	6.59	$\pm$	0.32	\\
050318	&	1.44	&	51.85	$\pm$	0.05	&	$47^{+15}_{-8}  $  	&	2.88	$\pm$	0.20	\\
050401	&	2.9	&	53.19	$\pm$	0.05	&	$118^{+18}_{-18}  $  	&	4.61	$\pm$	0.23	\\
050406	&	2.44	&	51.32	$\pm$	0.11	&	$25^{+35}_{-13}  $  	&	1.92	$\pm$	0.18	\\
050416	&	0.65	&	50.99	$\pm$	0.07	&	$15^{+2.3}_{-2.7}  $  	&	0.58	$\pm$	0.07	\\
050502	&	3.79	&	52.79	$\pm$	0.12	&	$93^{+55}_{-35}  $  	&	4.60	$\pm$	0.28	\\
050505	&	4.27	&	52.79	$\pm$	0.07	&	$70^{+140}_{-24}  $  	&	9.02	$\pm$	0.41	\\
050525	&	0.61	&	51.90	$\pm$	0.01	&	$81^{+1.4}_{-1.4}  $  	&	2.24	$\pm$	0.11\tablenotemark{d}	\\
050603	&	2.82	&	53.84	$\pm$	0.03	&	$344^{+52}_{-52}  $  	&	1.47	$\pm$	0.12	\\
050820	&	2.61	&	52.28	$\pm$	0.07	&	$246^{+76}_{-40}  $  	&	6.46	$\pm$	0.41	\\
050904	&	6.29	&	53.08	$\pm$	0.06	&	$436^{+200}_{-90}  $  	&	59.46	$\pm$	2.97\tablenotemark{d}	\\
050908	&	3.35	&	52.02	$\pm$	0.07	&	$41^{+9}_{-5}  $  	&	4.86	$\pm$	0.14	\\
050922	&	2.2	&	52.88	$\pm$	0.02	&	$198^{+38}_{-22}  $  	&	1.15	$\pm$	0.05	\\
051022	&	0.8	&	52.53	$\pm$	0.03	&	$510^{+22}_{-20}  $  	&	10.30	$\pm$	0.23	\\
051109	&	2.35	&	52.54	$\pm$	0.05	&	$161^{+130}_{-35}  $  	&	3.78	$\pm$	0.35	\\
060108	&	2.03	&	51.54	$\pm$	0.43	&	$65^{+600}_{-10}  $  	&	3.20	$\pm$	0.14	\\
060115	&	3.53	&	52.21	$\pm$	0.05	&	$62^{+19}_{-6}  $  	&	15.42	$\pm$	0.54	\\
060116	&	6.6	&	53.03	$\pm$	0.24	&	$139^{+400}_{-36}  $  	&	21.76	$\pm$	1.15	\\
060124	&	2.3	&	52.66	$\pm$	0.05	&	$237^{+76}_{-51}  $  	&	5.12	$\pm$	0.18	\\
060206	&	4.05	&	52.86	$\pm$	0.02	&	$75^{+12}_{-12}  $  	&	1.86	$\pm$	0.09	\\
060210	&	3.91	&	52.93	$\pm$	0.02	&	$149^{+400}_{-35}  $  	&	23.62	$\pm$	1.00	\\
060223	&	4.41	&	52.64	$\pm$	0.08	&	$71^{+100}_{-10}  $  	&	2.82	$\pm$	0.18	\\
060418	&	1.49	&	52.35	$\pm$	0.02	&	$230^{+[20]}_{-[20]}  $  	&	12.48	$\pm$	0.59	\\
060502	&	1.51	&	51.75	$\pm$	0.19	&	$156^{+400}_{-33}  $  	&	7.42	$\pm$	0.28	\\
060510	&	4.9	&	52.42	$\pm$	0.07	&	$95^{+[60]}_{-[30]}  $  	&	70.21	$\pm$	1.63	\\
060526	&	3.21	&	52.36	$\pm$	0.06	&	$25^{+[5]}_{-[5]}  $  	&	9.54	$\pm$	1.03	\\
060604	&	2.68	&	51.75	$\pm$	0.08	&	$40^{+[5]}_{-[5]}  $  	&	9.28	$\pm$	0.63	\\
060605	&	3.8	&	52.25	$\pm$	0.19	&	$169^{+[30]}_{-[30]}  $  	&	8.83	$\pm$	0.22	\\
060607	&	3.08	&	52.36	$\pm$	0.13	&	$120^{+190}_{-17}  $  	&	13.12	$\pm$	0.54	\\
\enddata
\label{ZLandEp}
\tablenotetext{a}{These values were obtained from Schaefer (2007); all appropriate references are located in that paper.}
\tablenotetext{b}{These values were calculated from light curves from BATSE, HETE, {\it Swift} and {\it Konus} websites.}
\tablenotetext{c}{These bursts were used by Firmani et al. (2006) to obtain the Firmani relation.}
\tablenotetext{d}{Indicates where an estimate of 5$\%$ error was used.}
\tablenotetext{e}{Indicates where an estimate of 10$\%$ error was used.}
\tablenotetext{f}{Values in square brackets indicate an estimate based on typical values (see Schaefer 2007).}
\end{deluxetable}

\begin{deluxetable}{c|ccc}
\tabletypesize{\scriptsize}
\tablecolumns{3}
\tablewidth{0pc}
\tablecaption{Expanding the Firmani Relation.}
\tablehead{\colhead{Relation} & \colhead{$\chi^2_r$} & \colhead{RMS} & \colhead{$\sigma_{sys}$}}
\startdata
19 Bursts, Firmani's Data & 0.74 & 0.14 & 0.00 \\
19 Bursts, Independent Data & 14.50 & 0.35 & 0.34 \\
60 Bursts, Independent Data & 15.89 & 0.41 & 0.38 \\
\\
\enddata
\label{Firmani}
\end{deluxetable}

\begin{deluxetable}{ccc}
\tabletypesize{\scriptsize}
\tablecolumns{3}
\tablewidth{0pc}
\tablecaption{RMS and Systematic Errors Values For Durations.}
\tablehead{\colhead{Duration Definition} & \colhead{$RMS$} & \colhead{$\sigma_{sys}$}}
\startdata
$T_{0.15}$	&	0.41	&	0.37	\\
$T_{0.30}$	&	0.40	&	0.36	\\
$T_{0.45}$ (Firmani Relation)	&	0.41	&	0.38	\\
$T_{0.50}$	&	0.41	&	0.37	\\
$T_{0.60}$	&	0.41	&	0.37	\\
$T_{0.75}$	&	0.41	&	0.38	\\
$\mathcal{T}_{15}$	&	0.46	&	0.39	\\
$\mathcal{T}_{30}$	&	0.41	&	0.38	\\
$\mathcal{T}_{45}$	&	0.37	&	0.33	\\
$\mathcal{T}_{50}$	&	0.38	&	0.38	\\
$\mathcal{T}_{60}$	&	0.41	&	0.36	\\
$\mathcal{T}_{75}$	&	0.40	&	0.36	\\
$S_{bolo}/P_{bolo}$	&	0.56	&	0.49	\\
$T_{90}$	&	0.46	&	0.41	\\
$T_{50}$	&	0.47	&	0.43	\\
$10 s$ ($E_{p} -L_{iso}$)	&	0.50	&	0.46	\\
$T_{0.15}/N_{peak}$	&	0.33	&	0.29	\\
$T_{0.30}/N_{peak}$	&	0.35	&	0.30	\\
$T_{0.45}/N_{peak}$	&	0.34	&	0.30	\\
$T_{0.50}/N_{peak}$	&	0.34	&	0.30	\\
$T_{0.60}/N_{peak}$	&	0.35	&	0.31	\\
$T_{0.75}N_{peak}$	&	0.35	&	0.31	\\
$\mathcal{T}_{15}/N_{peak}$	&	0.37	&	0.30	\\
$\mathcal{T}_{30}/N_{peak}$	&	0.31	&	0.27	\\
$\mathcal{T}_{45}/N_{peak}$	&	0.32	&	0.28	\\
$\mathcal{T}_{50}/N_{peak}$	&	0.35	&	0.30	\\
$\mathcal{T}_{60}/N_{peak}$	&	0.38	&	0.34	\\
$\mathcal{T}_{75}/N_{peak}$	&	0.38	&	0.35	\\
$S_{bolo}/P_{bolo}/N_{peak}$	&	0.41	&	0.33	\\
$T_{90}/N_{peak}$	&	0.42	&	0.38	\\
$T_{50}/N_{peak}$	&	0.41	&	0.37	\\
$10 s /N_{peak}$	&	0.52	&	0.51	\\
\enddata
\label{RMS}
\end{deluxetable}


\begin{thebibliography}{}


\bibitem[]{} Amati, L. et al. 2002, A\&A, 390, 81.
\bibitem[]{} Firmani, C., Ghisellini, G., Avila-Reese, V., \& Ghirlanda, G. 2006, MNRAS, 370, 185.
\bibitem[]{} Guidorzi, C., Frontera, F., Montanari, E., Rossi, F., Amati, L., Gomboc, A., Hurley, K., \& Mundell, C. G. 2005, MNRAS, 363, 315.
\bibitem[]{} Jimenez, R., Band, D., Piran, T., 2001, ApJ, 561, 171.
\bibitem[]{} Koshut, T. M., Paciesas, W. S., Kouveliotou, C., van Paradijs, J., Pendleton, G. N., Fishman, G. J., \& Meegan, C. A. 1996, ApJ, 463, 570.
\bibitem[]{} Norris, J. P., Bonnell, J. T., Nemiroff, R. J., Scargle, J. D., Kouveliotou, C., Paciesas, W. S., Meegan, C. A., Fishman, G. J. 1995, ApJ, 439, 542.
\bibitem[]{} Reichart, D. E., et al. 2007, MNRAS, 379, 619.
\bibitem[]{} Rizzuto, D. et al. 2001, ApJ, 552, 57.
\bibitem[]{} Rossi, F. et al. 2008, preprint, astro-ph/0802.0471
\bibitem[]{} Schaefer, B.E. 2003, ApJLett, 583, L71.
\bibitem[]{} Schaefer, B.E. 2004, ApJ, 602, 306.
\bibitem[]{} Schaefer, B.E. 2007, ApJ, 660, 16.
\bibitem[]{} Schaefer, B.E., Collazzi, A.C., 2007, 656, L53.

\end{thebibliography}
\end{document}